\documentclass[conference,letterpaper,10pt]{IEEEtran}
\IEEEoverridecommandlockouts

\usepackage[T1]{fontenc}
\usepackage{newtxtext}
\usepackage{cite}
\usepackage{amsmath,amssymb,amsfonts}
\usepackage[cmintegrals]{newtxmath}
\usepackage{array,textcomp,xcolor}

\usepackage[linesnumbered,ruled,vlined]{algorithm2e}
\usepackage{tcolorbox}
\usepackage{makecell}
\usepackage[caption=false,font=footnotesize]{subfig}
\usepackage{bbding}
\usepackage{multirow,booktabs}
\usepackage{listings}
\usepackage{pifont}
\usepackage{graphicx}
\usepackage{enumitem}
\usepackage{tikz}
\usepackage[hidelinks]{hyperref}
\providecommand{\Description}[1]{}
\let\oldding\ding

\DeclareFixedFont{\ttb}{T1}{lmtt}{bx}{n}{9.0} %
\DeclareFixedFont{\ttm}{T1}{lmtt}{m}{n}{9.0}  %
\definecolor{deepblue}{rgb}{0,0,0.5}
\definecolor{deepred}{rgb}{0.6,0,0}
\definecolor{deepgreen}{rgb}{0,0.5,0}
\definecolor{backcolour}{rgb}{0.95, 0.95, 0.96}

\newcommand\pythonstyle{\lstset{
language=Python,
basicstyle=\ttm,
backgroundcolor=\color{backcolour},
morekeywords={self},              %
keywordstyle=\ttb\color{deepblue},
emph={@rtai, submit_mpi, submit_rpc, rtai, wait, submit},          %
emphstyle=\ttb\color{deepred},    %
stringstyle=\color{deepgreen},
frame=tb,                         %
showstringspaces=false,
captionpos=b
}}

\lstnewenvironment{python}[1][]
{
\pythonstyle
\lstset{#1}
}
{}
\newcommand\pythoninline[1]{{\pythonstyle\lstinline!#1!}}
\renewcommand{\ding}[2][1]{\scalebox{#1}{\oldding{#2}}}

\author{\IEEEauthorblockN{Xiao Liu\textsuperscript{1,2}, Yujia Fu\textsuperscript{1},
    Yue Yu\textsuperscript{2}, Dan
    Huang\textsuperscript{1,*}\thanks{\textsuperscript{*}Corresponding author.}}
\IEEEauthorblockA{\textsuperscript{1}\textit{Sun Yat-sen University}, Guangzhou, China\quad
\textsuperscript{2}\textit{Peng Cheng Laboratory}, Shenzhen, China\\
\{liux276, fuyj25\}@mail2.sysu.edu.cn, yuy@pcl.ac.cn,
huangd79@mail.sysu.edu.cn}}

\begin{document}

\title{PairSight: Characterizing and Predicting Operator-Level
  Interference in GPU Sharing for DNN Inference}

\maketitle

\begin{abstract}
  Serving DNN inference is costly, yet a single task seldom saturates both the
compute units and the memory bandwidth of a GPU. GPU sharing raises
utilization by co-locating several tasks on one GPU, at the cost of
interference among them. Existing systems estimate this interference with
binary compute--memory rules, standalone scaling curves, or model-level
predictors. However, these estimates rest on a limited understanding of
interference. They leave the software and hardware properties of co-located
operators largely unexplored and give no slowdown prediction that is both
interpretable and accurate.
We conduct an operator-level study of interference under SM partitioning and use hardware
counters to trace it to contention on the shared memory path, where the
slowdown depends jointly on both co-located operators. Based on these
findings, we propose PairSight, a hardware-aware interference predictor
that learns how the hardware behaviors of co-located tasks interact and
predicts the latency of each task before co-location. On six DNN models
including LLMs, PairSight predicts latency under sharing with $4.3\%$ mean
error, against $19.0\%$ for a compute--memory intensity rule, and adapts to
unseen models and inputs with a few samples.

\end{abstract}


\section{Introduction}
\label{sec:intro}

\begin{figure*}[t]
    \centering
    \subfloat[Linear]{%
        \includegraphics[height=1.28in]{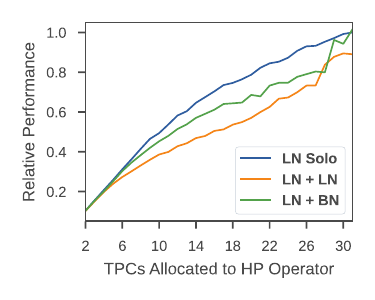}%
        \label{fig:ln}}%
    \hfill
    \subfloat[BatchNorm]{%
        \includegraphics[height=1.28in]{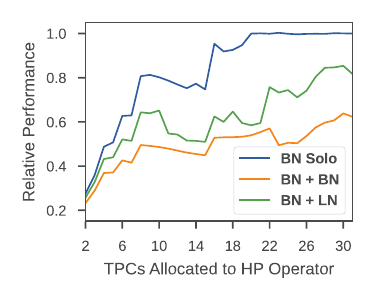}%
        \label{fig:bn}}%
    \hfill
    \subfloat[Pairwise HP slowdown]{%
        \includegraphics[width=0.22\textwidth,height=1.28in,keepaspectratio]{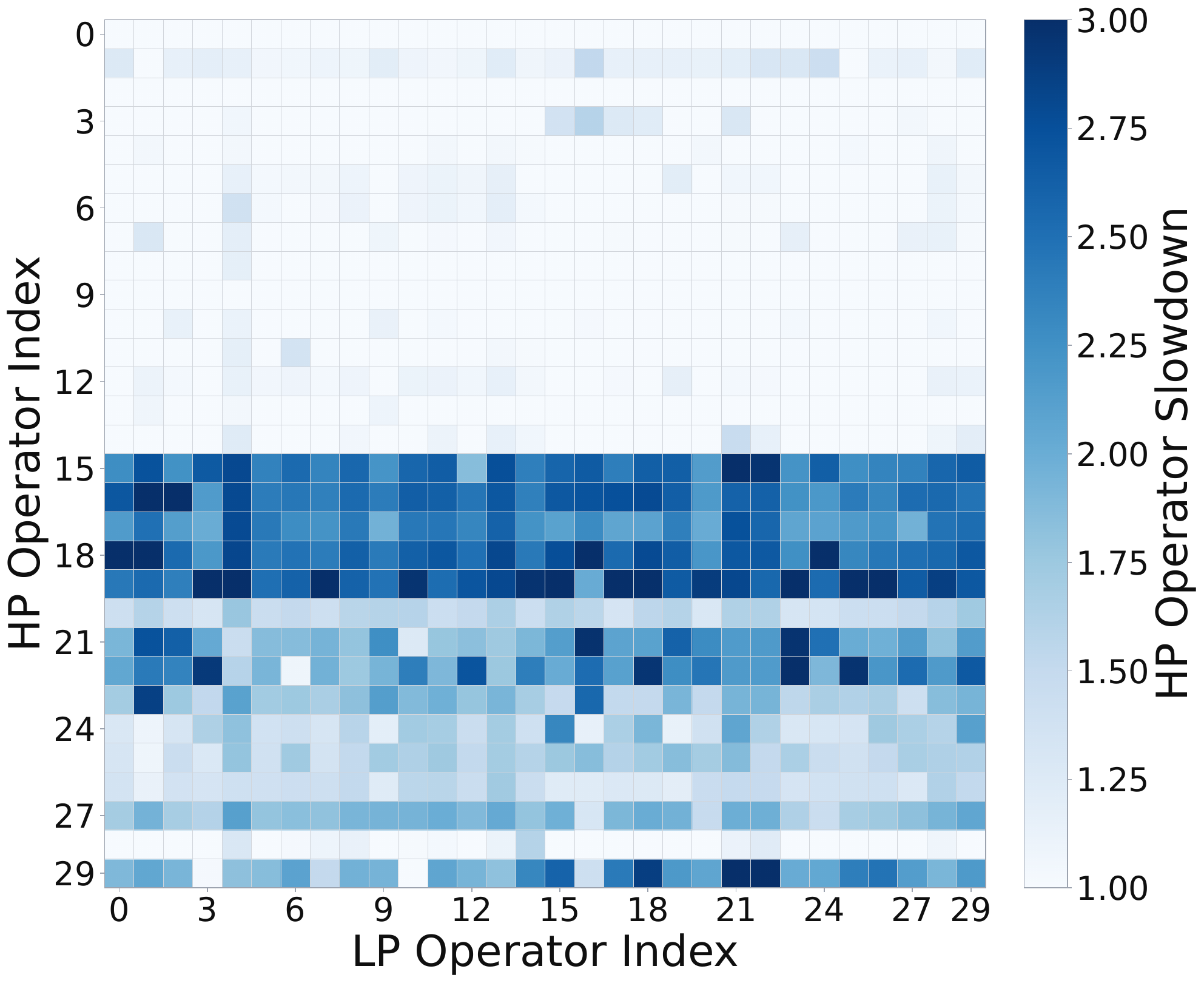}%
        \label{fig:heatmap}}%
    \hfill
    \subfloat[Fixed SM allocation, varying co-runner]{%
        \includegraphics[height=1.28in]{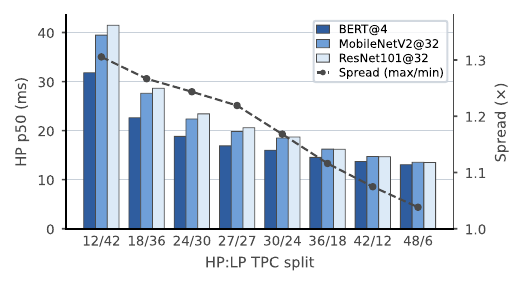}%
        \label{fig:corunner}}%
    \vspace{-0.1cm}
    \caption{Operator-level interference. (a, b) Normalized performance of a
        compute-intensive Linear and a memory-intensive BatchNorm operator as
        HP, alone and beside each co-runner, on an A10. (c) $\delta$ of
        operator pairs from four models on an A100, with HP operators as rows
        and LP operators as columns sorted by decreasing compute intensity.
        (d) Latency of ResNet-50 beside three co-runners at fixed SM
        allocations on an A100, with their spread dashed.}
    \label{fig:study}
    \vspace{-0.5cm}
\end{figure*}

Serving DNN inference is expensive and occupies large numbers of GPUs, yet a
single inference task seldom saturates both the compute units and the
memory bandwidth of a
GPU~\cite{kwon2023vllm,patel2024splitwise,strati2024orion}. GPU sharing
reclaims the unused resources by co-locating several tasks on one GPU and
dividing its time or its streaming multiprocessors (SMs) among them~\cite{chen2016baymax,yu2020salus,wang2023tgs,ng2023paella,dhakal2020gslice,choi2022serving,kim2022paris,li2022miso,xu2023igniter,chow2023krisp,zhang2025sgdrc,han2022microsecond,strati2024orion,shubha2024usher,zhang2025improving},
and whether it pays off depends on how much the co-located tasks slow each
other down.

Existing systems estimate this interference with a binary compute--memory
complementarity rule~\cite{strati2024orion,shubha2024usher}, standalone
scaling curves~\cite{dhakal2020gslice,choi2022serving,kim2022paris,chow2023krisp},
or predictors built per model on
identity~\cite{xu2023igniter,chen2025mudi,cui2021enable}.
However, these estimates reflect a limited understanding of interference
and do not relate the GPU performance metrics of individual operators to
their slowdown under co-location. They can neither identify which co-located
operators interfere nor quantify how much each operator slows down. For example, the
complementarity rule expects a memory-intensive operator to suffer little
interference beside a compute-intensive one, yet memory-intensive operators
slow down markedly beside compute-intensive and memory-intensive operators
alike.

We study interference in the setting adopted by recent GPU sharing systems,
where a latency-critical high-priority (HP) inference task shares a GPU with
a low-priority (LP) best-effort
task~\cite{han2022microsecond,strati2024orion,zhang2025sgdrc,chen2025mudi},
consistent with how production clusters co-locate latency-sensitive and
best-effort work~\cite{verma2015borg}. A scheduler in
this setting co-locates segments of consecutive operators from the two tasks
under an SM split. To keep the HP task within its latency budget, it must
predict how long each segment will run beside the other before co-location.
An underestimate violates the budget, and an overestimate rejects a feasible
co-location and forgoes LP throughput.

Predicting this latency is hard. First, SM partitions leave the L2 cache, the
DRAM channels, and the interconnect shared~\cite{strati2025gpusharing},
whereas MIG isolates them only across a few statically configured
instances~\cite{mig}. Second, the effect of the co-located task, the
co-runner, depends on the target and is hidden in aggregate statistics. Third, the prediction must precede co-location, yet measuring each of
the many candidate segments and SM splits on the GPU takes about $0.2$\,s, and
offline profiling of co-executions covers only short segments of known
models.

Our operator-level study shows that interference is an interaction between
the hardware behavior of two operators rather than a property of either task
or of its intensity. The memory traffic of the co-runner delays the accesses of an operator to
the shared L2 and DRAM path. The resulting slowdown is large for an
operator with heavy memory traffic and poor latency hiding, and small
otherwise. The memory traffic of both operators and the latency hiding of
each can be measured when each task runs alone, and PairSight, as hardware-aware interference
predictor, learns their interaction to predict latency before co-location.
Our contributions are as follows.
\begin{itemize}[leftmargin=*,nosep]
    \item \textbf{Interference study.} We characterize operator-level
          interference under SM partitioning, analyze its correlation with
          the hardware counters of both co-located operators, and model it as
          contention on the shared L2 and DRAM path. Interference is
          directional, operator-specific within the same intensity, and
          dependent on the co-runner at a fixed SM allocation.
    \item \textbf{PairSight.} We build PairSight on a gradient-boosted
          segment model that reads the standalone latency and memory-path
          counters of every operator of both tasks and predicts the latency
          of each task under a given SM split before co-location. Window
          composition extends it to segments longer than those in training,
          and continued training on a few samples adapts it to unseen models
          and batch sizes.
    \item \textbf{Evaluation.} We evaluate PairSight on 730k co-execution
          samples of six models including Llama-3.2 on an A100, covering
          accuracy, cost, ablation, and generalization. PairSight predicts the HP
          latency with $4.3\%$ mean error, against $19.0\%$ for an intensity
          rule.
\end{itemize}

\section{Interference among Concurrent DNN Tasks}
\label{sec:study}

\subsection{Shared Resources under SM Partitioning}
\label{sec:study_setting}

A DNN model executes as a sequence of operators, each usually one GPU
kernel. We partition the SMs between two co-located tasks at the granularity
of a Texture Processing Cluster (TPC), a pair of SMs, by attaching a TPC mask
to every kernel launch through \texttt{libsmctrl}~\cite{bakita2023hardware}.
Partitioning isolates registers, shared memory, the L1 cache, and the warp
schedulers, while the L2 cache, the DRAM channels, and the interconnect
remain shared~\cite{strati2025gpusharing}.

For a target operator, or a segment of consecutive operators, that runs on
$p$ TPCs beside a co-runner on $q$ TPCs, we define its slowdown as
\begin{equation}
    \delta = T_{\mathrm{share}}(p,q) / T_{\mathrm{solo}}(p),
    \label{eq:slowdown}
\end{equation}
where $T_{\mathrm{solo}}(p)$ is its standalone latency on the same $p$ TPCs.
The denominator removes the effect of receiving fewer SMs, and $\delta$
retains only the cost of concurrent execution.

\subsection{Observations}
\label{sec:study_obs}
\textbf{Directional interference.} Interference is directional, and a
memory-intensive operator slows down markedly more than a compute-intensive
one beside co-runners of either intensity. Figs.~\ref{fig:ln} and~\ref{fig:bn}
co-execute a compute-intensive \texttt{Linear} and a memory-intensive
\texttt{BatchNorm} operator, each as HP, beside each of the two operators.
In the complementary pair, Linear retains most of its standalone
performance. BatchNorm loses less beside Linear than beside another
BatchNorm, yet it remains far below its standalone performance.
A complementarity rule that admits this pair as low-interference
underestimates the slowdown of BatchNorm.

\textbf{Variation within the same intensity.} The slowdown varies widely
among operators of the same intensity.
Fig.~\ref{fig:heatmap} extends the measurement to operator pairs from four
real models. Compute-intensive HP operators stay near $\delta=1$ beside every
co-runner, whereas memory-intensive HP operators slow down markedly beside
both compute-intensive and memory-intensive co-runners. Even within the
memory-intensive group, one operator remains almost unaffected, and the
slowdown of a single HP operator ranges from negligible to more than
$2\times$ across its co-runners.

\textbf{Dependence on the co-runner.} The co-runner changes the slowdown of
a target even when the SM allocation
of the target is fixed. Fig.~\ref{fig:corunner} fixes the HP model and its
SM allocation and varies only the co-runner. With a small SM allocation,
ResNet-50 runs up to $1.31\times$ longer beside ResNet-101 than beside BERT,
and the spread narrows as the allocation grows. This spread bounds the error of any estimator
that ignores the co-runner.

\subsection{Where Interference Comes From}
\label{sec:study_model}

Under SM partitioning, interference arises mainly on the shared memory
path. The latency of a target operator can be approximated as
\begin{equation}
    T \approx T_{\mathrm{compute}} + N_{\mathrm{miss}}
    \left(L_{\mathrm{base}} + Q_{\mathrm{shared}}\right),
    \label{eq:memory_path}
\end{equation}
where $N_{\mathrm{miss}}$ counts the accesses of the target to the shared L2
and DRAM path, $L_{\mathrm{base}}$ is their uncontended service latency, and
$Q_{\mathrm{shared}}$ is the queueing delay added by concurrent traffic. The
co-runner can only raise $Q_{\mathrm{shared}}$, and the resulting cost
scales with $N_{\mathrm{miss}}$ and with the fraction of the added latency
that the target fails to hide. The model is consistent with the three
observations. A compute-intensive operator hides most of its memory
latency, whereas a memory-intensive operator pays the added queueing delay
on its memory accesses, which yields the directional interference.
Operators of the same intensity still differ in $N_{\mathrm{miss}}$ and in
how much latency they hide. Finally, $Q_{\mathrm{shared}}$ grows with the memory traffic of the co-runner, the
overlap of the two operators, and the SMs allocated to each, which makes
the slowdown depend on the co-runner even at a fixed SM allocation.

\begin{figure}[t]
    \centering
    \includegraphics[width=0.97\columnwidth]{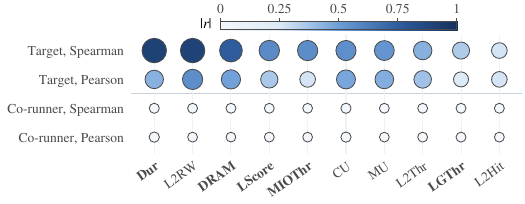}
    \vspace{-0.2cm}
    \caption{Correlation of hardware counters with the slowdown of the
        target. Bubble area and color encode $|r|$. All target correlations
        are significant at $p<0.05$, and no co-runner correlation is. Bold
        labels mark the counters used by PairSight. Labels denote standalone
        duration (Dur), L2 and DRAM bytes (L2RW, DRAM), long-scoreboard,
        MIO-throttle, and LG-throttle stalls (LScore, MIOThr, LGThr), compute
        and memory utilization (CU, MU), and L2 throughput and hit rate
        (L2Thr, L2Hit).}
    \label{fig:correlation}
    \vspace{-0.5cm}
\end{figure}

A correlation analysis of hardware counters identifies the shared L2 and
DRAM path as the main source of interference. Over 7.7k co-executed operator
pairs, we compute the correlation between $\delta$ of the target and ten
Nsight Compute~\cite{ncu_2025} counters of each operator
(Fig.~\ref{fig:correlation}). Every target counter is significantly
correlated with $\delta$, with $|r|$ above $0.9$ for standalone duration and
L2 traffic, whereas no co-runner counter is. The variation of $\delta$ is
thus dominated by differences between targets. Within groups that share the
target segment and the SM split, however, $\delta$ increases with the DRAM
traffic rate of the co-runner in two thirds of the 4.2k groups, which
indicates that the effect of the co-runner depends on the target. PairSight
takes standalone duration, DRAM traffic, and three memory-related stall
counters of both tasks as input and models their interaction with a
nonlinear learner.

\section{PairSight: Hardware-Aware Interference Prediction}
\label{sec:design}

\begin{figure}[t]
    \centering
    \includegraphics[width=0.96\columnwidth]{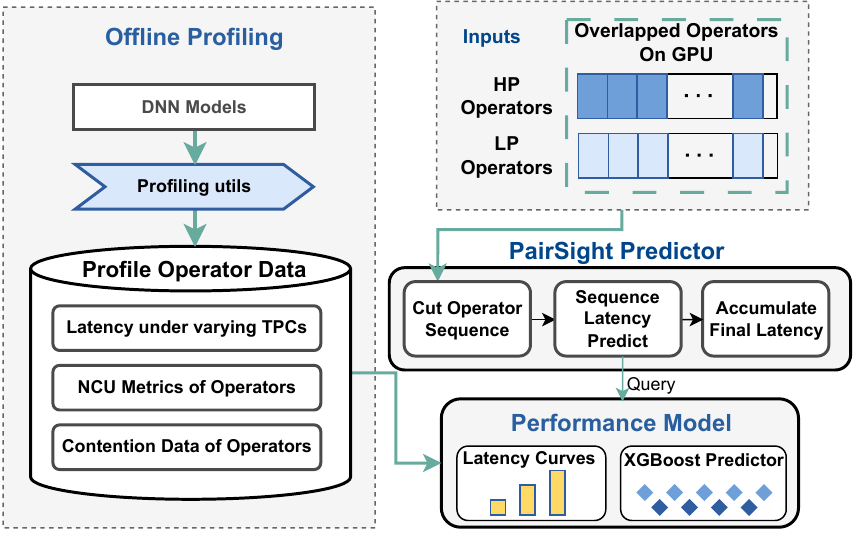}
    \caption{Overview of PairSight.}
    \label{fig:overview}
    \vspace{-0.6cm}
\end{figure}

PairSight predicts the co-execution latency of a candidate pair of segments
under a given SM split (Fig.~\ref{fig:overview}). Its core is
a segment model of two XGBoost regressors trained on measured co-executions
of segments with at most $16$ operators per side. The representation holds
one slot per operator, which limits the segment model to this length, and
PairSight serves longer segments by dividing both segments into windows
and composing the window predictions. Its inputs come from profiling each
DNN model once with Nsight Compute~\cite{ncu_2025} for per-operator counters
and once per TPC count for per-operator standalone latency.

The segment model predicts the slowdown of a target segment beside a
co-runner segment. For a target segment $a$ of consecutive operators on $p$
TPCs and a co-runner segment $b$ on $q$ TPCs, both within the trained
length, it returns
\begin{equation}
    \hat{\delta}_a = f\left(\Phi_a, \Phi_b, p, q\right), \qquad
    \hat{T}_a = T_{\mathrm{solo},a}(p)\,\hat{\delta}_a,
    \label{eq:predictor}
\end{equation}
where $\Phi_a$ and $\Phi_b$ describe the two segments and
$T_{\mathrm{solo},a}(p)$ comes from the per-TPC profile. Separate
regressors serve the HP and the LP side to match the directional
interference, and each regressor sees both segments.

The representation describes both segments operator by operator with the
quantities of Eq.~\eqref{eq:memory_path}. Each operator becomes one token
with its standalone duration at the assigned SM allocation, four memory-path
counters, and a presence flag. Total DRAM bytes represent
$N_{\mathrm{miss}}$. Long-scoreboard stalls measure how much memory latency
the operator fails to hide, and LG-throttle and MIO-throttle stalls, which
count warps blocked by full load-store and memory-I/O queues, indicate how
close the operator runs to the regime where $Q_{\mathrm{shared}}$ grows. For the co-runner, the same
counters measure the pressure that it adds to the shared path. Separate
tokens preserve which operator suffers unhidden memory latency and which
operator generates traffic, information that a per-segment sum would merge. The
input also carries $p$, $q$, and the standalone latency of both segments,
from which the model can weigh the fixed dispatch cost that every
co-execution adds to $\delta$.

The segment model consists of gradient-boosted regression trees, which
combine the two profiles without a hand-specified interaction form. We fit
them with XGBoost~\cite{chen2016xgboost} on $\log\delta$. By branching on
the unhidden memory latency of the target, the trees can route compute-bound segments to
leaves where co-runner counters barely matter and memory-bound segments to
leaves where these counters dominate, capturing the interaction of
Section~\ref{sec:study_model}.
Tree depth and learning rate are chosen on the validation set, and each
regressor trains on CPU with early stopping within $45$\,s. Adding trees to
an existing ensemble adapts the segment model to a new DNN model or batch
size with a few samples.

For segments beyond the trained length, PairSight composes predictions of
the segment model over windows. Both segments are divided in order into windows of at most
$16$ operators, and the $k$-th window $a_k$ of the target is queried beside
the $k$-th window of the co-runner under the same SM split, with the
standalone latency of each window as input. With $K$ windows on the side
with fewer operators, the target latency is
\begin{equation}
    \hat{T}_a = \sum_{k \le K} T_{\mathrm{solo},a_k}(p)\,\hat{\delta}_{a_k}
              + \sum_{k > K} T_{\mathrm{solo},a_k}(p).
    \label{eq:compose}
\end{equation}
Both segments start together, and windows beyond $K$ run after the shorter
segment has finished and add their standalone latency. Pairing windows by
index assumes that the two streams progress at comparable rates per
window, and
Section~\ref{sec:eval_generalization} measures the error of this
approximation on segments of up to $96$ operators.

\section{Evaluation}
\label{sec:eval}

\subsection{Experiment Setup}
\label{sec:eval_setup}

\textbf{Platform.} Experiments run on a server with eight NVIDIA
A100-SXM4-40GB GPUs connected over PCIe without NVLink and an Intel Xeon
Gold 6348 CPU.

\textbf{Models and Dataset.} We use ResNet-50, ResNet-101, MobileNetV2,
BERT-base, and Llama-3.2-1B and -3B~\cite{grattafiori2024llama3}, combined
into $20$ ordered HP/LP pairs of models and batch sizes. Each profiling
experiment co-executes an HP segment with an LP segment under a setting of
SM splits and repeats 30 runs. The co-execution latency $T_{\mathrm{share}}$
is measured with CUDA events on the stream of each task, whereas the standalone latency $T_{\mathrm{solo}}(p)$ sums the
device times of the operators. The 730k samples from 52k
experiments are divided by experiment into training, validation, and test
sets in an $80/10/10$ ratio.

\textbf{Baselines.} The baselines represent existing interference
estimates and separate the contribution of the learner from that of the
input. Following the compute--memory rule of Orion and
Usher~\cite{strati2024orion,shubha2024usher}, the \emph{intensity rule}
scales the standalone latency of each operator by a fixed factor per
intensity, and the \emph{complementarity table} uses the median training
slowdown per pair of intensities and SM split. \emph{Abacus}~\cite{cui2021enable},
a model-level predictor, feeds model identity, batch size, operator
positions, the SM split, and the standalone latencies to a multilayer
perceptron. \emph{Linear} is a ridge regression on the PairSight input, as
in linear contention models~\cite{han2024kace,choi2022serving}, and
\emph{PairSight-MLP} replaces the trees with a three-layer perceptron. All
learned methods predict $\log\delta$.

\begin{table}[t]
    \centering
    \caption{Prediction error and cost on the test set. HP and LP give the
        latency MAPE of each side, ${\le}10\%$ the share of HP predictions
        within $10\%$, and Llama the HP MAPE on the Llama pairs. Training uses
        $64$ CPU threads, and query time is per candidate in batches of 1k on
        $16$ cores.}
    \label{tab:baselines}
    \small
    \setlength{\tabcolsep}{2.4pt}
    \begin{tabular}{lrrrrrr}
        \toprule
        Method           & HP          & ${\le}10\%$ & LP          & Llama       & Train       & Query       \\
                         & (\%)        & (\%)        & (\%)        & (\%)        & (s)         & ($\mu$s)    \\
        \midrule
        Intensity rule   & $19.0$         & $37.9$          & $15.6$         & $45.0$         & --          & $0.4$       \\
        Compl.\ table    & $14.8$         & $50.0$          & $12.5$         & $39.7$         & $0.5$       & $0.4$       \\
        Linear           & $12.2$         & $50.9$          & $10.2$         & $24.0$         & $8$         & $1.1$       \\
        Abacus           & $9.2$          & $80.1$          & $8.0$          & $13.0$          & $111$       & $1.3$       \\
        PairSight-MLP    & $\mathbf{4.3}$ & $90.0$          & $\mathbf{2.4}$ & $\mathbf{8.3}$ & $135$       & $0.8$       \\
        \textbf{PairSight} & $\mathbf{4.3}$ & $\mathbf{90.6}$ & $\mathbf{2.4}$ & $10.3$         & $79$        & $5.3$       \\
        \bottomrule
    \end{tabular}
    \vspace{-0.6cm}
\end{table}

\subsection{Accuracy and Cost}
\label{sec:eval_baselines}

PairSight and PairSight-MLP are the most accurate predictors
(Table~\ref{tab:baselines}). PairSight errs by $4.3\%$ on the HP side and
$2.4\%$ on the LP side. The two take the same inputs, and differ only in the learner, so the accuracy stems from this operator-level input of both tasks rather than from a particular nonlinear learner.

\begin{figure}[t]
    \centering
    \vspace{0.12cm}
    \includegraphics[width=0.9\columnwidth]{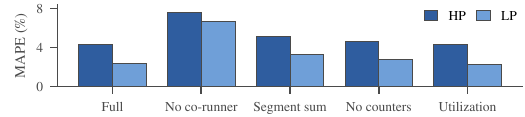}
    \vspace{-0.3cm}
    \caption{Design ablation. HP and LP latency MAPE with one design change
        at a time.}
    \label{fig:ablation}
    \vspace{-0.6cm}
\end{figure}

\subsection{Design Ablation}
\label{sec:eval_ablation}

Each ablation changes one design choice of PairSight at a time
(Fig.~\ref{fig:ablation}).
Removing the co-runner tokens produces the largest loss and raises the HP error
from $4.3\%$ to $7.7\%$ and the LP error from $2.4\%$ to $6.7\%$. Summing the
tokens of each segment raises the HP error to $5.1\%$, and keeping only the
operator durations of both segments raises it to $4.6\%$. Replacing the DRAM and stall counters with the compute and memory
utilization of each operator keeps the error at $4.3\%$, indicating that
continuous per-operator utilization of both tasks is as informative as the
memory-path counters.

\begin{table}[t]
    \centering
    \caption{Generalization to unseen models and batch sizes. HP and LP
        latency MAPE (\%) on held-out groups for predictors trained on the
        four classic models at default batch sizes. The bottom rows continue
        training PairSight on $k$ new experiments per held-out pair.}
    \label{tab:generalization}
    \small
    \setlength{\tabcolsep}{3.2pt}
    \begin{tabular}{lrrrrrr}
        \toprule
                         & \multicolumn{2}{c}{Llama} & \multicolumn{2}{c}{Batch 16} & \multicolumn{2}{c}{Batch 64} \\
        \cmidrule(lr){2-3}\cmidrule(lr){4-5}\cmidrule(lr){6-7}
        Method           & HP       & LP          & HP       & LP          & HP       & LP          \\
        \midrule
        Intensity rule   & $45.0$   & $27.7$      & $47.1$   & $39.3$      & $16.8$   & $18.8$      \\
        Linear           & $37.7$   & $43.9$      & $36.9$   & $31.1$      & $19.6$   & $18.6$      \\
        Abacus           & $65.9$   & ${>}100$    & $35.7$   & ${>}100$    & $19.3$   & ${>}100$    \\
        PairSight-MLP    & $25.5$   & $25.8$      & $33.9$   & $28.6$      & $10.0$   & $8.3$       \\
        PairSight        & $30.2$   & $18.5$      & $35.1$   & $26.8$      & $15.0$   & $9.9$       \\
        \midrule
        \multicolumn{7}{l}{\emph{PairSight, continued training}} \\
        $k=20$           & $17.4$   & $11.3$      & $24.6$   & $16.9$      & $13.9$   & $8.1$       \\
        $k=100$          & $13.5$   & $8.0$       & $18.9$   & $10.9$      & $10.4$   & $6.3$       \\
        All              & $9.1$    & $5.1$       & $12.0$   & $7.6$       & $6.8$    & $3.7$       \\
        \bottomrule
    \end{tabular}
    \vspace{-0.5cm}
\end{table}

\subsection{Generalization and Adaptation}
\label{sec:eval_generalization}

PairSight transfers to unseen DNN models and batch sizes better than the
existing estimates and adapts to them with a few samples (Table~\ref{tab:generalization}).
Trained only on the four classic models at their default batch sizes and
tested on the Llama pairs and on unseen batch sizes $16$
and $64$, PairSight and PairSight-MLP give the two lowest errors in every
held-out group, whereas Abacus, which relies on model identity, errs by more
than $100\%$ on the LP side.  Continued training on $100$ new data samples lowers the HP error from $30.2\%$ to $13.5\%$ on Llama, with each update taking at most
$5$\,s on CPU.

Window composition lets PairSight predict segments longer than its
$16$-operator window without new samples. On $400$ separate
co-executions with segments of $24$ to $96$ operators and SM splits absent
from training, PairSight errs by $5.5\%$ and $5.4\%$ on the HP and LP
sides, against $9.8\%$ and $17.0\%$ for the intensity rule.
On the $80$ asymmetric co-executions, where the segment with fewer windows
runs longer and the windows that Eq.~\eqref{eq:compose} pairs by index do
not coincide in time, it errs by $4.9\%$ and $5.5\%$.

\section{Conclusion}
\label{sec:conclusion}

Interference among co-located DNN tasks under SM partitioning arises mainly
on the shared L2 and DRAM path and depends jointly on both operators.
PairSight learns this interaction from the per-operator standalone profiles
of both tasks and predicts the latency of each task with $4.3\%$ mean error,
against $19.0\%$ for an intensity rule. Continued training on a few samples
and window composition extend it to unseen models, batch sizes, and longer segments.
Accurate interference prediction before co-location lets GPU sharing
systems select co-locations that meet their performance requirements.

\clearpage
\typeout{PAIRSIGHT-REFERENCES-START-PAGE=\thepage}

\end{document}